\documentclass[reprint,amsmath,amssymb,aps,twocolumn,floatfix]{revtex4}

\usepackage{graphicx}
\usepackage{dcolumn}
\usepackage{bm}
\usepackage{leftidx}
\usepackage{txfonts}
\usepackage{txfonts}
\usepackage{physics}
 
 \usepackage{amsthm}
\usepackage{enumitem}
\usepackage{appendix}
\usepackage{color}
\usepackage{xcolor}

\DeclareUnicodeCharacter{2212}{-}

\begin{document}

\preprint{APS/123-QED}

\title{Self-regularized entropy: What does black hole entropy predict for tests of Kerr no-hair theorem?
}

\author{Shokoufe Faraji}
\email{s3faraji@uwaterloo.ca}
\affiliation{Department of Physics and Astronomy, University of Waterloo,Waterloo, Canada}
\affiliation{Perimeter Institute for Theoretical Physics,Waterloo, Canada}
\affiliation{Waterloo Centre for Astrophysics, University of Waterloo,Waterloo, Canada}

\author{Niayesh Afshordi}
\email{nafshordi@pitp.ca}
\affiliation{Department of Physics and Astronomy, University of Waterloo,Waterloo, Canada}
\affiliation{Perimeter Institute for Theoretical Physics,Waterloo, Canada}
\affiliation{Waterloo Centre for Astrophysics, University of Waterloo,Waterloo, Canada}

\begin{abstract}

We compute the canonical, or brick-wall, entropy of a massless scalar field in a quantum black hole model whose strong field exterior is described phenomenologically by the static $q$-metric, also known as the Zipoy-Voorhees metric. This geometry is an exact vacuum deformation of Schwarzschild with a small quadrupolar distortion parameter, $q$. Using WKB counting of trapped near horizon cavity modes, we show that this deformation changes the near horizon density of states so that the usual Schwarzschild brick-wall ultraviolet divergence is self-regularized, eliminating the need for an ad hoc proper distance cutoff within the perturbative regime studied here. Treating the Hawking temperature and Bekenstein-Hawking entropy of a Schwarzschild black hole of the same mass as external thermodynamic benchmarks, we obtain an analytic entropy-motivated deformation scale, $|q|\sim 0.2$, across the stellar-to-supermassive black hole mass range. Through a stationary extension, this scale maps phenomenologically onto percent-to-tens-of-percent violations of the Kerr multipole relations, providing observational targets for ngEHT imaging, LISA extreme mass ratio inspirals, and third generation ground based gravitational wave tests.
\end{abstract}

\maketitle

\section{Introduction}\label{sec:intro}

Despite the success of black hole thermodynamics, the microscopic origin of the Bekenstein-Hawking entropy remains unsettled; a variety of horizon degrees of freedom have been proposed e.g.,
\cite{PhysRevD.7.2333,1998PhRvL..81.4293H,1996PhLB..379...99S,1997NuPhS..57...65I,2000gr.qc.....5126A,2003PhRvL..90h1301D,2004CQGra..21.5245M,2006CQGra..23.3849D,Saravani:2012is,2021PhRvD.104l1502C,2023PhLB..84137901O,2023arXiv230404491N}, yet no single picture is universally accepted. In this work, we explore a conservative, geometry-based option: we ask whether a minimal deformation of the near-horizon geometry can render the brick-wall mode count finite, thereby providing a controlled arena for horizon scale microphysics. Concretely, we focus on the high frequency near-horizon cavity (trapped) modes that dominate the brick-wall entropy in echo capable, horizonless compact objects, and we distinguish them from classical black hole quasinormal modes defined by purely ingoing conditions at the horizon and outgoing behaviour at infinity.

The brick-wall model of ’t Hooft \cite{tHooft:1984kcu,tHooft:1996rdg} remains the simplest way to estimate the canonical (thermodynamic) entropy of quantum fields around a black hole, but in the exactly spherical Schwarzschild geometry the mode sum diverges logarithmically and one is forced to introduce an ad‑hoc proper distance cutoff. In this paper, we show that the brick-wall ultraviolet divergence is self-regularized once the Schwarzschild exterior is perturbed by a sufficiently small but nonzero quadrupolar deformation.

An implication of a brick-wall model for a quantum black hole is that we can send and receive signals from the quantum structure near the horizon, within a finite time. This has motivated searches for ``echoes'' following black hole mergers in gravitational wave e.g, \cite{Abedi:2016hgu,Cardoso:2016rao,Oshita:2019sat,Wang:2019rcf,Oshita:2023tlm,Vellucci:2022hpl,Arrechea:2024nlp} (see \cite{Abedi:2020ujo,Berti:2025hly} for reviews). However, this would also imply that a stationary spacetime may not obey the no-hair theorem, as it can learn about the absence of a regular horizon in a finite time (e.g., \cite{Bah:2021jno}). In this paper, we use the $q$-metric as an analytic vacuum laboratory for this purpose: it introduces a single controlled quadrupolar deformation while keeping the exterior governed by the Einstein equations.

Our goal is to isolate the single geometric ingredient that makes the Schwarzschild brick-wall sum diverge and to test how a minimal, physically motivated deformation changes the near horizon density of states. The static Zipoy-Voorhees $q$-metric \cite{osti_4201189,PhysRevD.39.2904} is ideal for this. It is an exact, asymptotically flat, one parameter deformation of Schwarzschild that remains a vacuum solution of Einstein’s equations and reduces smoothly to Schwarzschild as $q\,\to\,0$. The single parameter
$q$ encodes a quadrupolar departure from spherical symmetry; the first nontrivial \textit{hair} allowed once perfect symmetry is relaxed. Working in vacuum general relativity avoids modeling ambiguities from matter sources or
modified field equations and keeps all changes traceable to geometry alone.

We work with the static $q$-metric line element \cite{osti_4201189,PhysRevD.39.2904}, a one-parameter vacuum deformation of Schwarzschild, and restrict to first order in $|q|\ll 1$. Using the brick-wall framework, we perform a WKB count of the high-frequency trapped modes in the near-horizon cavity and compute the associated canonical entropy in closed form. We find that the Schwarzschild logarithmic divergence is removed: for sufficiently small nonzero $|q|$ in the regime considered here, the mode sum converges without introducing an ad hoc proper-distance cutoff. Treating the Hawking temperature and the Bekenstein-Hawking entropy of a Schwarzschild black hole of the same mass as external thermodynamic benchmarks, we obtain an entropy-motivated deformation scale of order $|q|\sim 0.2$ for astrophysical black holes. In a stationary extension of the $q$-metric, the same parameter induces Kerr-violating shifts in the lowest multipoles (captured by the Kerr-null combinations $Q_1$ and $Q_2$), making this scale an observational target for ngEHT and LISA.

For nonzero $q$, the static $q$-metric is horizonless and contains a singularity at the would-be horizon, so it is not interpreted here as a literal classical black hole exterior. Instead, it is used only as an effective description of the strong field exterior region of a quantum black hole model, with the microscopic completion of the near-horizon region left unspecified. In particular, the asymptotic Hawking temperature and the Bekenstein-Hawking entropy for a Schwarzschild black hole of the same mass are treated as external thermodynamic inputs. Within this framework, the role of the $q$-metric is to modify the near-horizon redshift and transverse geometry in order to regulate the WKB density of states responsible for thermodynamic entropy, in a manner consistent with the Einstein field equations. We do not claim that astrophysical objects are exact $q$-metrics. Rather, the $q$-metric provides the simplest vacuum setting that isolates a single quadrupolar deformation and allows one to track its consequences analytically; (a) preserves the approximate universal Rindler\,$\times$\,transverse near horizon structure, (b) introduces precisely one controlled deformation that governs the ultraviolet behaviour of the mode sum, and (c) connects directly to observable multipoles. More general bumpy or rotating spacetimes introduce additional scales that may change quantitative details. However, they are not expected to alter the qualitative self-regularization mechanism introduced here and are thus left for future work.

\smallskip
The rest of the manuscript is organized as follows: Section \ref{sec:qmetric} reviews the $q$-metric. In Section \ref{sec:wave} we derive the scalar wave operator and the exact radial Klein-Gordon problem, while Section \ref{sec:counting} performs the WKB counting and presents the entropy. Section  \ref{sec:obs} offers observational prospects for measuring quadrupole, and Section \ref{sec:conclusion} gathers the main implications. In this work, we adopt the convention $G = c = \hbar = k_B = 1$.

\section{Space-time} \label{sec:qmetric}

The $q$-metric describes static, axially symmetric, and asymptotically flat exact solutions to the {\it vacuum} Einstein field equations. By incorporating the quadrupole moment, the $q$-metric accounts for the deformation of the gravitational field which is not captured by the spherically symmetric Schwarzschild solution.  Consequently, the $q$-metric characterizes the exterior gravitational field of an isolated static axisymmetric object. The presence of a quadrupole can change the geometric properties of spacetime and affect the timelike and lightlike trajectories e.g., \cite{2021A&A...654A.100F,2025PhRvD.111d5003A,2025A&A...699A.266F,2025EPJC...85..148F}. As a result, the $q$-metric is a particularly useful toy model for the gravitational fields of realistic astrophysical bodies where the effects of higher-order multipole moments cannot be ignored. The metric in Schwarzschild-like coordinates $(t, x, y, \phi)$ is presented as follows \cite{osti_4201189,PhysRevD.39.2904}
\begin{align}
   ds^2 = &\left(1-\frac{2m}{r}\right)^{1+{ q}} dt^2 - \left(1-\frac{2m}{r}\right)^
    {-{ q}} \nonumber\\
    &\left[ \left(1+\frac{m^2\sin^2\theta}{r^2-2mr}\right)^{-{ q}(2+{q})}\left(\frac{dr^2}{1-\frac{2m}{r}}+r^2d\theta^2\right)\right. \nonumber\\
  &\left.+r^2\sin^2\theta d\phi^2\right],\
\end{align}
\label{qmetric} 
where $t \in (-\infty, +\infty)$, $r \in (2m, +\infty)$, $\theta \in [0,\pi]$, and $\phi \in [0, 2\pi)$, also $m$ is a parameter which is expressed in the dimension of length. This metric contains two free parameters, namely $m$, and quadrupole parameter $q$. In the case of vanishing $q$, the Schwarzschild metric is recovered. Furthermore, values of $q > 0$ describe the exterior field of an oblate central object, while $q < 0$ is related to a prolate source. Additionally, the $q$-metric, aside from a central curvature singularity at $r=0$, has another singularity located at a finite distance from the origin at $r=2m$ for any chosen nonzero value of quadrupole. By Geroch's definition \citep{1970JMP....11.2580G}, for multipole moment and for avoiding a negative mass distribution, the quadrupole parameter is restricted to the domain of $q\in(-1,\infty)$.

\subsection{Echo time and entropy bound}\label{subsec:echot}

The time scale for a null signal to reach the Schwarzschild radius $r=2m$ from the light ring $r=3m$  and back, in the static $q$-metric is given by
\begin{eqnarray}
    t_{\rm echo} &&\simeq 2\int_{2m}^{3m} dr \left(1-\frac{2m}{r}\right)^{-1-q} \left(\frac{r^2-2mr+m^2\sin^2\theta }{r^2-2mr}\right)^{-\frac{q(2+q)}{2}}\nonumber\\
    &&\simeq 8 m q^{-2} +{\cal O}(m q^{-1}) = \frac{1}{\pi T_{\rm H} q^2}.\label{eq:techo}
\end{eqnarray}
The $\theta$-dependence in the integrand produces only an $\mathcal{O}(m)$ modulation and does not affect the leading $m q^{-2}$ scaling %
\footnote{To first order in $q$, the light-ring radius shifts to
$r_{\rm LR}=m(3+2q)$; using $3m$ as the upper limit changes $t_{\rm echo}$ only at $\mathcal{O}(q^{0})$, hence subleading to the $q^{-2}$ term.}. We take the asymptotic Hawking temperature to be that of a
Schwarzschild black hole with the same mass, $T_{\rm H} \simeq 1/(8\pi m)$, as an external thermodynamic input. 
Motivated by the conjectured correspondence between the fast scrambling \cite{Sekino:2008he} and echo time scales \cite{Saraswat:2019npa,Oshita:2023tlm}, we adopt the following
conservative thermodynamic consistency hypothesis: the entropy associated with the trapped
near-horizon thermal atmosphere 
should not parametrically exceed the
Bekenstein-Hawking entropy of a Schwarzschild black hole of the same mass. Following \cite{Oshita:2023tlm}, the echo delay time of a finite-entropy compact object with approximately Rindler near-surface geometry,
admits an upper bound (the ``quantum limit of silence'') of the form $t_{\rm echo}\lesssim \ln S/(2\pi T_{\rm H})$.
Applying this bound to the $q$-metric echo time \eqref{eq:techo} yields

\begin{equation} \label{eq:bound}
 q^2 \gtrsim \frac{2}{\ln S}. 
\end{equation}
Equation (\ref{eq:bound}) should be read as a conservative rough 
estimate for a minimal $|q|$, required to avoid overcounting entropy.

In what follows (Sections \ref{sec:wave}-\ref{sec:obs}), we compute the canonical entropy directly in the $q$-metric by near-horizon WKB mode-counting, which yields a sharper relation $q=q(S)$ for this spacetime and thus a more accurate bound or estimate on $q$.


\subsection{Small-$q \ll 1$ near horizon limit}

 In this part, we derive the near horizon line element and the associated tortoise coordinate for the static $q$-metric \eqref{qmetric}. Throughout, we keep the leading non-vanishing power of the affine distance to the horizon and work in $|q|\ll1$ limit. This computation will prepare us for the WKB mode counting in the next section, which shall yield a finite enetropy of Hawking radiation in the q-metric spacetime. 

In the limit $ q \ll 1 $, the metric simplifies and the exponents in $\rho$ become approximately symmetric. Introducing the dimensionless parameter $\hat{\rho}=1-2m/r$, so that $\hat{\rho}\to 0^+$ as $r\to 2m$. We then pass to the Gaussian-normal coordinate $\rho$ by
\begin{equation}\label{eq:first-beta-f}
\rho = \hat{C} \, \hat{\rho}^{\,\beta}, \quad \text{where} \quad \beta :=\frac{q^2}{1+q+q^2},
\end{equation}
with an arbitrary constant scale $\hat{C}$ (irrelevant hereafter). In these coordinates, this 2D $(t,\rho)$ sector takes a Rindler-like form $-f(\rho)dt^2+d\rho^2$, with 
\begin{equation}
f(\rho) = \left(\frac{\rho}{4m}\right)^{\alpha}, \qquad \alpha:=2(1-\beta)=\frac{2(1+q)}{1+q+q^{2}},
\end{equation}
approaching Rindler in $q \to 0 \Rightarrow \alpha \to 2$ limit. The angular 2-metric is a mild deformation of the round sphere. To leading order in $|q|$ the full near horizon line element is
\begin{align}\label{eq:nh_metric}
 ds^2 &= f(\rho) \left( -dt^2 + \frac{1}{f(\rho)} d\rho^2 \right)\nonumber\\
&+ 4m^2 \left[\left(\frac{\rho}{4m}\right)^{2q} d\theta^2
+ \left(\frac{\rho}{4m}\right)^{-2q} \sin^2\theta \, d\phi^2\right].   
\end{align}
This procedure is equivalent to zooming into a local Rindler
patch, but it preserves the $q$‑dependent exponents that control the ultraviolet behaviour of the mode sum . 

We can consider spacetime in equation (\ref{eq:nh_metric}) as a generalized Rindler patch, times a transverse 2-manifold. We are using this Rindler form of metric to simplify the computation of entropy in the WKB approximation. In the high frequency limit, a field mode is a localized wave packet whose phase advances according to the null Hamilton-Jacobi relation $g^{\mu\nu}p_{\mu}p_{\nu}=0$. The number of modes with frequency below $\omega$ is, to leading order, the phase‑space volume on a constant time slice divided by
$(2\pi)^{3}$: $N(\omega)\ \sim\ \frac{1}{(2\pi)^{3}}\,
\int d^{3}x\,d^{3}p\,\Theta\,\bigl[\omega-\mathcal{H}\,(x,p)\bigr]$ where $\mathcal{H}$ is determined by the local metric. Therefore, only two local
inputs control the ultraviolet behavior:
(a) the redshift $-g_{tt} = f(\rho)$ in the radial Rindler sector, and (b) the intrinsic metric of the transverse 2-surface (which fixes the map between angular quantum numbers and transverse momenta). For near horizon Schwarzschild, $-g_{tt}\propto\rho^{2}$ and the transverse radius
$r(\rho)$ tends to a constant, $2m$. The local phase space measure near the horizon contains $d\rho/\rho$ - a logarithmically divergent integral- since the redshift creates arbitrarily many radial ``boxes'' of size set by the local wavelength. In the static $q$-metric, the transverse circumference is not constant as in equation \eqref{eq:nh_metric}. This single geometric
change modifies the phase space measure by a factor $\rho^{\,\pm 2q}$ coming from the
transverse area and the centrifugal barrier. 
These two ingredients are precisely what the deformed Rindler $\times$ (deformed $S^{2}$) neighborhood encodes, and they are exactly what turn the divergent Schwarzschild log into a convergent power when $|q|>0$ as we see in the rest of the paper. 

\subsubsection{Tortoise Coordinates}
\label{subsec:tortoise}

We define the tortoise coordinate $r^{*}$ in the usual way, so that the radial part of the Klein-Gordon operator has the Schrödinger form, and integrating gives

\begin{align}
dr_{*}&=\frac{d\rho}{\sqrt{f(\rho)}}
      =(4m)^{\alpha/2}\,\rho^{-\alpha/2}\,d\rho \nonumber\\
&\Rightarrow\quad
r_{*}=\frac{(4m)^{\alpha/2}}{\beta}\,\rho^{\,\beta},
\label{eq:tortoise}
\end{align}
where $\beta$ is given by equation \eqref{eq:first-beta-f}. Inverting $\rho=\left(\frac{\beta\,r_{*}}{(4m)^{\alpha/2}}\right)^{\,1/\beta}$, turns the metric \eqref{eq:nh_metric} to
\begin{align}\label{eq:metric_rstar}
ds^{2}&=
-(4m)^{-\alpha}\left(\frac{\beta\,r_{*}}{(4m)^{\alpha/2}}\right)^{\,\alpha/\beta}\,dt^{2}
+(4m)^{-\alpha}\left(\frac{\beta\,r_{*}}{(4m)^{\alpha/2}}\right)^{\,\alpha/\beta}\,dr_{*}^{2} \nonumber\\
&\quad+4m^{2}\left[(4m)^{-2q}\left(\frac{\beta\,r_{*}}{(4m)^{\alpha/2}}\right)^{\,2q/\beta}d\theta^{2}\right. \nonumber \\
 &\quad\left.+(4m)^{\,2q}\left(\frac{\beta\,r_{*}}{(4m)^{\alpha/2}}\right)^{\,-2q/\beta}\sin^{2}\theta\,d\phi^{2}\right]\,.
\end{align}
Note that in the Schwarzschild limit $q\to 0$ one has $\alpha\to 2$ (equivalently $\beta\to 0$), so the near-horizon metric
\eqref{eq:nh_metric} reduces to the standard Rindler$\times S^2$ form
\begin{equation}
ds^2 \simeq -\left(\frac{\rho}{4m}\right)^2 dt^2 + d\rho^2 + 4m^2\left(d\theta^2+\sin^2\theta\,d\phi^2\right),
\end{equation}
with surface gravity $\kappa=1/(4m)$. Correspondingly, the tortoise coordinate
\begin{equation}
r_*=\int \frac{d\rho}{\sqrt{f(\rho)}}=\frac{4m}{\beta}\left(\frac{\rho}{4m}\right)^{\beta},
\end{equation}
approaches a logarithm as $\beta\to 0$:
\begin{equation}
r_*\xrightarrow[\beta\to 0]{} 4m\,\ln\,\left(\frac{\rho}{4m}\right)+\text{const},
\quad\text{so}\quad
\frac{dr_*}{d\rho}\xrightarrow[\beta\to 0]{}\frac{4m}{\rho}.
\end{equation}
This is the origin of the familiar near-horizon phase space factor $dr_*\propto d\rho/\rho$ (hence a logarithmic divergence) in the Schwarzschild brick-wall mode count. For $0<\beta\ll 1$ (i.e.\ sufficiently small but nonzero $|q|$ in our regime), one instead has $dr_*\propto \rho^{\beta-1}d\rho$, and the same near-horizon integral becomes a convergent power, $\int_0 d\rho\,\rho^{\beta-1}\sim \rho^{\beta}/\beta$.




\section{Wave Operator Setup}
\label{sec:wave}

For the massless scalar field the equation of motion is 
\begin{equation}
\Box\,\varphi \;=\;
\frac{1}{\sqrt{-g}}\,
\partial_\mu\,\bigl(\sqrt{-g}\,g^{\mu\nu}\partial_\nu\varphi\bigr)=0.
\label{eq:KG}
\end{equation}
Using the near horizon metric \eqref{eq:nh_metric} one obtains $
\sqrt{-g}=4m^{2}\sqrt{f}\sin\theta$. A component by component evaluation (see appendix \ref{app:KG}) of equation \eqref{eq:KG} gives

\begin{align}\label{eq:box-open}
\Box \varphi &\,=
-\frac{1}{f(\rho)}\,\partial_t^{2}\varphi
+\partial_{\rho}^{2}\varphi
+\frac{1}{2}\,\frac{f'(\rho)}{f(\rho)}\,\partial_{\rho}\varphi \\
&+\frac{1}{4m^{2}}\left[
\left(\frac{\rho}{4m}\right)^{-2q}
\frac{1}{\sin\theta}\,\partial_{\theta}\,\left(\sin\theta\,\partial_{\theta}\varphi\right)
+\left(\frac{\rho}{4m}\right)^{2q}
\frac{1}{\sin^{2}\theta}\,\partial_{\phi}^{2}\varphi
\right]\nonumber.
\end{align}


\subsection{Separation of variables} \label{subsec.sep}
To obtain a single radial equation we use the standard mode ansatz
\begin{equation}
\varphi(t,\rho,\theta,\phi)=e^{-i\omega t}\,Y_{\ell m}(\theta,\phi)\,\psi(\rho),
\quad
\Delta_{S^2}Y_{\ell m}=-\ell(\ell+1)Y_{\ell m}.
\end{equation}
Because the angular operator carries unequal $\rho^{\mp 2q}$ weights in the $\theta$ and $\phi$ parts, exact separation into a single ODE holds rigorously in the axisymmetric sector $m=0$. For general $m$ we proceed by projecting onto the standard $Y_{\ell m}$ basis and using a large-$\ell$ (high angular momentum) approximation appropriate to the ultraviolet mode count. In this regime the detailed angular eigenfunctions can change, but the leading high-eigenvalue density is controlled by the transverse area (Weyl asymptotics) and is therefore insensitive to $O(1)$ anisotropies. Accordingly, we replace the weighted angular operator by an effective $\ell(\ell+1)$ eigenvalue at the level needed to capture the near-horizon WKB scaling; this may affect overall numerical prefactors but does not affect the power-law exponents that control convergence of the brick-wall sum.\footnote{For each fixed $\rho$, the induced two-metric from \eqref{eq:nh_metric} has area element $4m^2\sin\theta\,d\theta\,d\phi$, independent of $\rho$ and $q$ (since $g_{\theta\theta}g_{\phi\phi}=(4m^2)^2\sin^2\theta$). By Weyl's law, the asymptotic counting of large angular eigenvalues depends only on this area, so the ultraviolet scaling of the angular density of states is unchanged up to $O(1)$ factors.}

Projecting \eqref{eq:box-open} onto $Y_{\ell m}$ and using this approximation yields the radial equation

\begin{equation}\label{eq:radial-rho}
\psi''(\rho)+\frac{1}{2}\frac{f'(\rho)}{f(\rho)}\,\psi'(\rho)
+\left[\frac{\omega^2}{f(\rho)}-\frac{\ell(\ell+1)}{r^2(\rho)}\right]\psi(\rho)=0,
\end{equation}
where
\begin{equation}
r^2(\rho)=4m^2\left(\frac{\rho}{4m}\right)^{2q}.
\end{equation}
To remove the first derivative term we pass to tortoise coordinates
\eqref{eq:tortoise}, $dr_{*}=d\rho/\sqrt{f(\rho)}$, and perform the Liouville transform $\Psi(\rho):=f(\rho)^{-1/4}\psi(\rho)$.
Writing $\Psi(r_*)=\Psi(\rho(r_*))$ we obtain the Schrödinger form

\begin{equation}
\frac{d^{2}\Psi}{dr_{*}^{2}}
+\bigl[\omega^{2}-V_{\ell}(\rho)\bigr]\Psi=0,  
\end{equation}
where
\begin{equation}
V_{\ell}(\rho)=f(\rho)\,
 \left[
   \frac{\ell(\ell+1)}{r^{2}(\rho)}
  +\frac14\,\left(\frac{f'(\rho)}{f(\rho)}\right)^{2}
  -\frac12\,\frac{f''(\rho)}{f(\rho)}
 \right].
\label{eq:Schro-rho}
\end{equation}
For our $f(\rho)=\bigl(\rho/(4m)\bigr)^{\alpha}$ one has $f'/f=\alpha/\rho$ and
$f''/f=\alpha(\alpha-1)/\rho^{2}$, hence simplifies the bracket
\begin{equation}
\frac14\Bigl(\frac{f'}{f}\Bigr)^{2}-\frac12\frac{f''}{f}
   =\frac{\alpha\beta}{2\,\rho^{2}},
\end{equation}
Therefore, we obtain
\begin{equation}
V_{\ell}(\rho)=
\frac{\ell(\ell+1)}{4m^{2}}
     \left(\frac{\rho}{4m}\right)^{\alpha-2q}
\,+\;
\frac{\alpha\beta}{32\,m^{2}}
     \left(\frac{\rho}{4m}\right)^{\alpha-2}\, .
\label{eq:Vofrho}
\end{equation}
\paragraph*{In tortoise coordinates.}
Using $\rho=\left(\frac{\beta\,r_{*}}{(4m)^{\alpha/2}}\right)^{\,1/\beta}$ and noting $\alpha-2=-2\beta$, the $f$-derivative (Liouville) term becomes a universal inverse-square wall,
\begin{equation}
\frac{\alpha\beta}{32\,m^{2}}
     \left(\frac{\rho}{4m}\right)^{\alpha-2}
=\frac{\alpha}{2\beta}\,\frac{1}{r_{*}^{2}},
\end{equation}
and the centrifugal term scales as
\begin{align}
\frac{\ell(\ell+1)}{4m^{2}}
     &\left(\frac{\rho}{4m}\right)^{\alpha-2q}=\frac{\ell(\ell+1)}{4m^{2}}
 \left(\frac{\beta\,r_{*}}{4m}\right)^{\,p},
 \end{align}
 where
\begin{align}
 p := \frac{\alpha-2q}{\beta}= \frac{2}{q^2}-2-2q.\label{eq:p}
\end{align}
 Thus, the master equation becomes
\begin{equation}\label{eq:Schrodinger-rstar}
\frac{d^{2}\Psi}{dr_{*}^{2}}
+\Bigl[\omega^{2}-V_{\ell}(r_{*})\Bigr]\Psi=0,
\end{equation}
where
\begin{equation}
V_{\ell}(r_{*})
   =\frac{\ell(\ell+1)}{4m^{2}}
     \left(\frac{\beta\,r_{*}}{4m}\right)^{\,p}
     +\frac{\alpha}{2\beta}\,\frac{1}{r_{*}^{2}}.
\end{equation}
Equation \eqref{eq:Schrodinger-rstar} is the precise radial master equation that enters the WKB mode counting.  All powers and coefficients follow directly from the metric exponents $\alpha$ and $\beta$.

It is worth emphasizing that the nature and boundary conditions used to define QNMs in exotic compact objects (ECO), radically differ from those of classical black holes, as such objects lack regular horizons (e.g., \cite{Cardoso:2016rao,Wang:2019rcf}). Unlike the black hole QNMs that are mainly defined by the centrifugal barrier, the ECO QNMs are generically trapped near the would-be horizon, due to an inner Dirichlet-type boundary condition. As such, we focus on the near-horizon cavity: the region between the inner brick-wall/singularity (modeling the onset of quantum microstructure) and the outer centrifugal barrier. In this finite interval, the modes behave as standing waves, and the WKB condition is the usual Bohr-Sommerfeld quantisation for a bounded Schrödinger problem. The resulting mode density agrees with the high-frequency / large-$\ell$ QNM density up to sub-leading $\mathcal{O}(1)$ shifts in the overtone numbers, which do not affect the ultraviolet behaviour of the brick-wall entropy.

It is worth emphasizing that classical black hole quasinormal modes (e.g., \cite{Cardoso:2016rao,Wang:2019rcf}) are defined by purely ingoing behaviour at the horizon and outgoing behaviour at infinity, whereas an ECO with an inner reflective (Dirichlet-type) boundary supports a near-horizon cavity whose resonant modes are trapped between the inner boundary (represented by the inner WKB turning point at the end of the near-horizon region, modeling the onset of quantum microstructure) and the outer centrifugal barrier. For our purposes we focus on the high-frequency standing-wave spectrum in this finite cavity, for which the WKB condition is the usual Bohr-Sommerfeld quantisation of a bounded Schr\"odinger problem. This approximation captures the leading large-$\ell$ / high-frequency mode density relevant for the ultraviolet behaviour of the brick-wall entropy; subleading $O(1)$ phase shifts and the small imaginary parts associated with leakage through the barrier do not affect the convergence properties of the mode sum.





\subsection{Classical turning points and cavity length}
\label{subsec:turning}

The WKB approximation is valid in the domain where the radial wave number $k_{r}^{2}\,=\,\omega^{2}-V_{\ell}(\rho)$ is positive, i.e.\ between two
classical turning points. For the potential $V_{\ell}(\rho)$ in equation \eqref{eq:Vofrho}, the outer turning point $\rho_{\max}$ is defined by
$k_{r}^{2}(\rho_{\max})=0$. The inner turning point $\rho_{\mathrm{in}}(\omega)$ is set by the inverse-square
term in the effective potential (discussed below) and approaches $\rho=0$ in the high-frequency regime relevant
for the ultraviolet mode count.

\paragraph{Outer turning point.}
Using $f(\rho)=\bigl(\rho/(4m)\bigr)^{\alpha}$ and
$r^{2}(\rho)=4m^{2}\bigl(\rho/(4m)\bigr)^{2q}$, the transverse wavenumber is
\begin{equation}
k_{\perp}^{2}(\rho):=\frac{\ell(\ell+1)}{r^{2}(\rho)}
          \simeq\frac{\ell^{2}}{4m^{2}}\left(\frac{\rho}{4m}\right)^{-2q},
\end{equation}
Setting $\omega^{2}=f(\rho_{\max})\,k_{\perp}^{2}(\rho_{\max})$; hence for $\ell\gg 1$ the inverse-square term in \eqref{eq:Vofrho} is subleading at the outer turning point and is neglected in determining $\rho_{\max}$. We obtain
\begin{align}
\left(\frac{\rho_{\max}}{4m}\right)^{\alpha-2q}
&=\frac{4m^{2}\,\omega^{2}}{\ell^{2}} \nonumber\\
&\Longrightarrow\quad
\rho_{\max}
=4m\left(\frac{4m^{2}\omega^{2}}{\ell^{2}}\right)^{\,1/(\alpha-2q)} .
\label{eq:rho_max}
\end{align}

\paragraph{Tortoise position of the turning point.}
With equation \eqref{eq:tortoise}, $r_{*}=\dfrac{(4m)^{\alpha/2}}{\beta}\,\rho^{\beta}
=\dfrac{4m}{\beta}\left(\dfrac{\rho}{4m}\right)^{\beta}$, we obtain
\begin{equation}
  r^{*}_{\max}
   =\frac{4m}{\beta}\,
     \left(\frac{4m^{2}\omega^{2}}{\ell^{2}}\right)^{\,\beta/(\alpha-2q)}, \left(\approx \frac{4m}{q^2}\left(\frac{2 m \omega}{\ell} \right)^{q^2}\right)
\label{equ:rstar_max}   
\end{equation}
which approaches $t_{\rm echo}/2$ (equation \ref{eq:techo}), in the $q \ll 1$ limit.

\paragraph{Inner turning point (why one may set $\rho_{\min}=0$).}
Besides the centrifugal term, the effective potential contains the inverse-square contribution generated by the $f$-derivative terms in equation \eqref{eq:Schro-rho},
\begin{equation}
V_{\mathrm{L}}(\rho)
= f(\rho)\,\frac{\alpha\beta}{2\rho^{2}}
= \frac{\alpha\beta}{32\,m^{2}}\left(\frac{\rho}{4m}\right)^{\alpha-2},
\quad 0<\alpha<2,\ \beta>0,
\end{equation}
where we used $f(\rho)=\bigl(\rho/(4m)\bigr)^{\alpha}$ and
$\alpha-2=-2\beta$. Since $\alpha<2$ for every $|q|<1$, one has
$V_{\mathrm{L}}\to+\infty$ as $\rho\to0$, so the inner (classical)
turning point is determined by
\begin{align}
\omega^{2}=V_{\mathrm{L}}(\rho_{\mathrm{in}})
&\;\;\Longrightarrow\;\;
\left(\frac{\rho_{\mathrm{in}}}{4m}\right)^{\alpha-2}
=\frac{32\,m^{2}\,\omega^{2}}{\alpha\beta}\nonumber \\
&\;\;\Longrightarrow\;\;
\rho_{\mathrm{in}}
=4m\left(\frac{\alpha\beta}{32\,m^{2}\,\omega^{2}}\right)^{\,1/(2-\alpha)}.
\end{align}
In tortoise coordinates $r^{*}=\dfrac{4m}{\beta}\left(\dfrac{\rho}{4m}\right)^{\beta}$, this maps to
\begin{equation*}
r^{*}_{\mathrm{min}}=\frac{4m}{\beta}
 \left(\frac{\rho_{\mathrm{in}}}{4m}\right)^{\beta}=\frac{4m}{\beta}
 \left(\frac{\alpha\beta}{32\,m^{2}\,\omega^{2}}\right)^{\,1/2}
= \frac{1}{\omega}\,\sqrt{\frac{\alpha}{2\beta}} \approx \frac{1}{\omega |q|} \,,
\end{equation*}
which tends to zero for large frequencies. Since $\omega\,r^{*}_{\mathrm{min}}=\sqrt{\alpha/(2\beta)}$ is independent of $\omega$, keeping a nonzero $r^{*}_{\mathrm{min}}$
only shifts the Bohr-Sommerfeld overtone number by an $\omega$-independent constant and therefore does not
affect the WKB density of states $\partial n/\partial\omega$, the scaling with $\omega/\ell$, or the ultraviolet convergence of the brick-wall entropy.
Accordingly, we may set $\rho_{\min}=0$ (equivalently $r^{*}_{\min}=0$) in \eqref{eq:cavity_length} at the level relevant for the ultraviolet mode count.

\paragraph{Cavity length.}
The proper WKB cavity length is
\begin{equation}
L(\omega,\ell)=r^{*}_{\max}-r^{*}_{\min}.
\label{eq:cavity_length}
\end{equation}
We just obtained $r^{*}_{\min}\to 0$ at high frequencies and, moreover, it only induces an $\omega$-independent shift in the overtone number, so for the ultraviolet scaling we may set $L(\omega,\ell)\simeq r^{*}_{\max}$.

For $q=0$ (Schwarzschild) equation \eqref{equ:rstar_max} reduces to the familiar logarithmic behavior $L\sim 4m\ln\big(\ell/(m\omega)\big)$+ const. For any $0<|q|\ll 1$, it becomes a power law in $\omega/\ell$ displayed by equation \eqref{equ:rstar_max}; this change of scaling underlies the self-regularisation of the mode sum discussed in the next section.

We note that our analytic treatment uses the near-horizon (Rindler$\times$transverse) form of the metric and is under control for sufficiently small $|q|$. This is adequate to determine the ultraviolet behaviour of the brick-wall entropy, since the Schwarzschild divergence originates entirely from the $\rho\to 0$ region.
Moreover, the modes that dominate the ultraviolet mode count have large angular momentum: the outer turning point $\rho_{\max}(\omega,\ell)$ decreases with increasing $\ell$ (equation \eqref{eq:rho_max}). Therefore, for $\ell\gg m\omega$ the WKB cavity lies deep in the near-horizon region where the approximation applies. Global deviations of the geometry at larger radii can modify finite $O(q)$ contributions and overall normalization, but they do not control the regularization mechanism, which is fixed by the near-horizon scaling exponents.

\section{Mode counting and canonical entropy}
\label{sec:counting}

In this section we derive the canonical entropy of a single massless scalar field from WKB counting of the near-horizon cavity modes, evaluated at the asymptotic temperature $T_{\rm H}$ adopted in Sec.~\ref{subsec:echot}.

\subsection{WKB radial spectrum}

For each $(\ell,m)$ the radial wave-number is $k_r(\rho)=\sqrt{\omega^2-V_{\ell}(\rho)}$.
Let $r_*^{\min}(\omega)$ and $r_*^{\max}(\omega,\ell)$ denote the classical turning points (Sec.~\ref{subsec:turning}).
In the WKB approximation, for fixed $(\ell,m)$ the radial overtone number $n:=N_\ell(\omega)$ is given by
\begin{align}\label{eq:n}
n=\frac{1}{\pi}\int_{r_*^{\min}}^{r_*^{\max}} k_r\,dr_* \, .
\end{align}
In the high-frequency regime relevant for the ultraviolet mode count, $k_r\simeq\omega$ throughout most of the cavity and the turning-point regions contribute only subleading $O(1)$ phase shifts. Moreover, $r_*^{\min}$ produces an $\omega$-independent shift in $n$ (Sec.~\ref{subsec:turning}), so for the purpose of the mode density we may approximate $n \simeq \omega L(\omega,\ell)/\pi$ with $L(\omega,\ell)\simeq r_*^{\max}$. Using equation \eqref{equ:rstar_max} we obtain
\begin{align}\label{eq:LLL}
L(\omega,\ell)\simeq \frac{4m}{\beta}\,
\left(\frac{4m^{2}\omega^{2}}{\ell^{2}}\right)^{\,\beta/(\alpha-2q)}.
\end{align}

Finally, writing $L(\omega,\ell)=C(q,m)\,\ell^{-s}\omega^{s}$ gives an expression for the mode frequency $\omega_{n\ell}$ in terms of the angular harmonic and overtone numbers, $\ell$ and $n$:
\begin{eqnarray}
    &&\omega_{n\ell} = \left(\pi \over C\right)^{\frac{1}{1+s}} n^{\frac{1}{1+s}} \ell^{\frac{s}{1+s}},\\
    &&s := \frac{2}{p}=\frac{2\beta}{\alpha-2q} = \frac{q^2}{1-q^2-q^3},\\
    &&C(q,m) := \frac{4m}{\beta}\,(4m^{2})^{\,\beta/(\alpha-2q)}=\frac{2(1+q^2+q^3)}{q^2}(4m^{2})^{\frac{1+s}{2}}.\nonumber\\
\end{eqnarray}

Note that $s>0$ as long as $q<0.75$.

\subsection{Radial density of states}

Equipped with the spectrum of the near-horizon trapped modes, we can compute their thermal entropy. 

The canonical entropy of a bosonic harmonic oscillator with frequency $\omega$ at temperature $T$ is given by 
\begin{equation}
    \mathcal{S}(w)=\dfrac{w}{e^{w}-1}-\ln\,\bigl(1-e^{-w}\bigr),
\end{equation}
in terms of its dimensionless frequency $w := \omega/T$. 

Using the large-$\ell$ angular-density approximation discussed in Sec.~\ref{subsec.sep}, we treat the sum over angular modes by the standard replacement $\sum_m\to(2\ell+1)$ and convert the $\ell$ sum to an integral, which is sufficient for the ultraviolet scaling. The total thermal entropy is then given by a discrete sum over all the modes:
\begin{align}\label{eq:entropy_sum}
S(q,T) &\equiv \sum^\infty_{\ell=0} (2\ell+1)\sum_{n=1}^{\infty}\mathcal{S}\left(\frac{\omega_{n\ell}}{T}\right) 
\simeq  \int_{0}^{\infty}2\ell d\ell \sum_{n=1}^{\infty}\mathcal{S}(w)\nonumber\\
&= \int_{0}^{\infty} \left( \sum_{n=1}^{\infty} 2\ell \frac{\partial\ell(w,n)}{\partial w} \right)\mathcal{S} (w) dw,
\end{align}
with $n=\left(\frac{C}{\pi}\right)\ell^{-s}\omega^{s+1}$, and differentiating at fixed $n$ gives
\begin{align}
\sum_{n=1}^{\infty} 2\ell\, \frac{\partial \ell}{\partial w}
= \frac{2(1+s)}{s} \, \left(\frac{C}{\pi}\right)^{\frac{2}{s}} \, \zeta\,\left(\frac{2}{s}\right)\, T^{\frac{2(1+s)}{s}}w^{\frac{2(1+s)}{s}-1}.
\end{align}
Therefore
\begin{align*}
S(q,T) =  \frac{2(1+s)}{s} \, \left(\frac{C}{\pi}\right)^{\frac{2}{s}} \, \zeta\,\left(\frac{2}{s}\right)\, 
 T^{\frac{2(1+s)}{s}} \int_{0}^{\infty}w^{\frac{2}{s} + 1}\, \mathcal{S}(w)\, dw,
\end{align*}
and using 
\begin{align*}
\int_{0}^{\infty} w^{u}\, \mathcal{S}(w)\, \mathrm{d}w
= (u+2)\, \Gamma(u+1)\, \zeta(u+2), \quad (u > -1)
\end{align*}
with $u=p+1$ and $p=\frac{2}{s}$ we obtain
\begin{align*}
S(q,T) = (p+2)\, \left(\frac{C}{\pi}\right)^{p}\, T^{p+2}\, \zeta(p)(p+3)\, \Gamma(p+2)\, \zeta(p+3).
\end{align*}

If we adopt the asymptotic Hawking temperature $T_{\rm H}=(8\pi m)^{-1}$ (Sec.~\ref{subsec:echot}), and $p=\frac{2}{s}$ we find

\begin{align*}
\left(\frac{C}{\pi}\right)^{p}\, T_{\rm H}^{p+2}=
(4 m^{2})\left(\frac{4 m}{\beta \pi}\right)^{p} (8 \pi m)^{-(p+2)}
&= 2^{-(p+4)}\, \beta^{-p}\, \pi^{-(2p+2)},\,
\end{align*}
the factors of $m$ cancel out. Thus, the final result is
\begin{equation}\label{eq:Sfinal}
S(q,T)=
\frac{(p+2)(p+3)}{2^{\,p+4}\,\pi^{\,2p+2}}\;
\beta^{-p}\;
\Gamma(p+2)\,\zeta(p)\,\zeta(p+3) \left(T \over T_{\rm H} \right)^{p+2},
\end{equation}
where $p \approx \frac{2}{q^2}$ and $\beta \approx q^2$ from equations (\ref{eq:p}) and (\ref{eq:first-beta-f}). This entropy is finite for all sufficiently small $|q|\neq0$. Note that in our thermodynamic input $T=T_H$, so the explicit factors of $T/T_H$ drop out and the final closed form result depends only on $q$; we keep $\tau$ up to last stage only to make the replacement $T\neq T_H$ transparent. Thus
\begin{equation}\label{eq:Sfinal2}
S(q,T)=
\frac{(p+2)(p+3)}{2^{\,p+4}\,\pi^{\,2p+2}}\;
\beta^{-p}\;
\Gamma(p+2)\,\zeta(p)\,\zeta(p+3).
\end{equation}
In contrast, as $q\to0$ one has $p\to\infty$ (hence $s(q)\to0^{+}$), the cavity length reverts to the Schwarzschild logarithm, and the entropy reproduces the usual brick wall divergence. In the formal zeta-regularized notation used earlier, this appears as the symbolic limit $\bigl[2^{-3}/q^{4}\bigr]\Gamma(0)\,\zeta(1)$ when $q\to 0$. In this limit, by using $\zeta(p+k) \to 1$, and Stirling’s formula for $\Gamma(p+2)$ we find:
\begin{equation}\label{eq:Sapprox}
    \ln S (q,T) = -\frac{2}{q^2}\ln\left(e \pi^2 q^4\times \frac{T}{T_{\rm H}}\right) +{\cal O}\left(1 \over q\right)
\end{equation}

\subsection{Discussion on the closed-form result Equation \eqref{eq:Sfinal}}

In this result, by self-regularization we mean the following kinematic statement: within the small-$|q|$ near-horizon geometry used in this work, the $q$-dependent anisotropy modifies the WKB density of trapped near-horizon cavity modes so that the Schwarzschild brick-wall logarithm is replaced by a convergent power law, without imposing an explicit proper distance cutoff by hand. In this sense the black hole ``shapes itself to keep its entropy finite'' \footnote{We thank the anonymous referee for the suggestion of this phrasing}, i.e., the same strong field geometry that defines the near-horizon cavity also fixes the local redshift and the transverse wave operator, and hence the phase space measure entering the WKB mode count. The parameter $q$ is then treated as an effective deformation parameter, with its scale inferred by matching the resulting canonical entropy $S(q)$ to the macroscopic benchmark $S_{\rm BH}$.

From the geometric standpoint, the parameter $q$ controls an anisotropic near-horizon deformation. Changing $q\to -q$ interchanges the roles of the angular directions in the wave operator (the $\rho^{\mp 2q}$ weights swap between the $\theta$ and $\phi$ sectors), corresponding to the opposite quadrupolar distortion (oblate vs.\ prolate). The ultraviolet entropy regularization mechanism is controlled by the near-horizon tortoise scaling: for $q=0$ one has the Schwarzschild logarithm $r_*\sim \ln\rho$, whereas for sufficiently small nonzero $|q|$ the tortoise coordinate becomes a power $r_*\propto \rho^{\beta}$ with $\beta>0$, so the near-horizon phase-space factor is a convergent power rather than a divergent log. Therefore, within the perturbative small-$|q|$ regime studied here, the canonical entropy remains finite for either sign of $q$, while the sign affects the orientation/sign of the associated quadrupolar multipole deviations (our Kerr null magnitudes $Q_1,Q_2$ depend on $|q|$ by construction in Subsection \ref{sec:multipole}).

Additionally, we note that we assume a KMS thermal state at infinity at the adopted asymptotic temperature $T_H$, with respect to the static Killing time $t$ used in the mode expansion. In the Schwarzschild limit this corresponds to the Hartle-Hawking thermal state (equilibrium). We do not consider the Boulware vacuum, which is empty at infinity, singular at the horizon in the black hole limit and does not describe Hawking emission. Importantly, the standard Schwarzschild brick-wall divergence persists even in a regular thermal state. The finiteness result here will arise from the $q$-dependent modification of the near-horizon density of states rather than from the choice of vacuum.


\section{Observational prospects} 
\label{sec:obs}

In what follows, we use $q$ purely as a phenomenological deformation parameter in the external multipole structure and lensing geometry, in the same spirit as other bumpy-metric parametrisations used in tests of the Kerr hypothesis; the forecasts are sensitivity estimates, not evidence that the exact static q-metric is realized in nature. For any four-dimensional, non-extremal black hole
\begin{equation}
 S_{\rm BH}= \frac{A_H}{4G\hbar}
 \sim 4\pi\,\Bigl(\frac{m}{m_{\mathrm P}}\Bigr)^{2},   
\end{equation}
where $m_{\mathrm P}=2.18\times10^{-5}\,\mathrm{g}$, is the Planck mass \footnote{For nonextremal Kerr, $8\pi M^{2}\le A_H\le 16\pi M^{2}$ in $G=c=\hbar=1$, so $\ln S_{\rm BH}$ changes by at most $\ln 2$, negligible for our purposes.}. Therefore, $\ln S_{\rm BH}$ grows only logarithmically with the mass. Using the echo-time bound \eqref{eq:bound} obtained in Sec.~\ref{eq:techo},
\begin{equation}
|q|\;\gtrsim\;\sqrt{\frac{2}{\ln S_{\rm BH}}}\;,
\end{equation}
we find that across the stellar-to-supermassive interval
$5\,M_\odot\le m\le10^{10}\,M_\odot$, corresponding to $S_{\rm BH}\sim10^{78}\text{--}10^{97}$, the bound varies only weakly, from $\sim 0.11$ at the stellar-mass end down to $\sim 0.095$ at the most massive.

However, using the closed-form canonical entropy equation \eqref{eq:Sapprox}, we can sharpen this estimate by matching $S(q,T)$ to the microscopic entropy scale $S_{\rm BH}$ and solving for the corresponding $q(S_{\rm BH})$ in the small-$|q|$ regime. 
 
For the astrophysically relevant small $|q|$ we can invert \eqref{eq:Sapprox} with the product-log (Lambert) function $W$, defined by $W(z)\,e^{W(z)}=z$ for $z>0$ to obtain
\begin{align}\label{eq:accurate-bound}
    q(S)\approx \sqrt{\frac{4W\left(\frac{\ln S_{\rm BH} \times \sqrt{T/T_{\rm H}}}{4\pi \sqrt{e}}\right)}{\ln S_{\rm BH}}}. 
\end{align}
Plugging $T \approx T_{\rm H}$ and $S_{\rm BH} \simeq 10^{78} - 10^{97}$ (i.e., $\ln S \simeq 180 - 223$) into the estimated formula above, we obtain the corresponding estimate
\begin{equation}\label{eq:qpred}
    |q|_{\rm new}\;\simeq\;0.18\,- \,0.19,
\end{equation}
which varies by less than 5\% across 10 orders of magnitude in mass, for the black hole entropies across the stellar to supermassive mass range. Similarly, changing the ratio of black hole temperature, $T$ to the standard Hawking temeparture $T_{\rm H}$ by up to a factor of 2, which may be expected due to quantum effects near the horizon (e.g., \cite{Oshita:2020dox}), only changes $q$ by less than 6\%. This weak dependence on entropy is useful: within this framework, the entropy-motivated deformation scale is essentially mass and temperature independent across the stellar-to-supermassive range, with $|q|\sim 0.2$. When translated (Sec.~\ref{sec:multipole}) into stationary multipole diagnostics, this corresponds to percent-to-tens-of-percent departures from the Kerr multipole relations, while the exterior remains close to Schwarzschild beyond a few gravitational radii in the sense of a small quadrupolar deformation.







\subsection{From static to stationary: Multipoles}\label{sec:multipole}  

The analysis in this paper has deliberately focused on the static $q$-metric, which is the only input needed to establish the near horizon self-regularisation of the canonical entropy. To connect the deformation parameter $q$ to stationary observables, solely to indicate how a nonzero $q$ would manifest as controlled deviations from the Kerr multipole hierarchy. We use a known stationary vacuum extension of the $q$metric \cite{Toktarbay_2014,Faraji_2022}. In prolate spheroidal coordinates $(t,x,y,\phi)$ it can be written in Lewis-Papapetrou form as
\begin{align}
ds^{2}=&-f\,(dt-\omega\,d\phi)^{2}
+\frac{\sigma^{2}}{f}\times\\
&\left[e^{2\gamma}(x^{2}-y^{2})\left(\frac{dx^{2}}{x^{2}-1}+\frac{dy^{2}}{1-y^{2}}\right)
+(x^{2}-1)(1-y^{2})\,d\phi^{2}
\right],\nonumber
\label{eq:stationary_qmetric}
\end{align}
with
\begin{align}
&f=\frac{A}{B},\\
&\omega=-2\left(a+\sigma\,\frac{C}{A}\right),\\
&e^{2\gamma}=\frac14\left(1+\frac{m}{\sigma}\right)^{2}\,
\frac{A}{(x^{2}-1)^{1+q}}\left(\frac{x^{2}-1}{x^{2}-y^{2}}\right)^{(1+q)^{2}} .
\label{eq:stationary_qmetric_functions}
\end{align}
The auxiliary functions $A,B,C,a_\pm,b_\pm,\lambda,\eta$ are collected in
Appendix~\ref{app:stationary_qmetric} for completeness. The parameters $(m,a,q)$ correspond to the mass monopole, angular momentum dipole, and an independent quadrupolar deformation, respectively, with $\sigma^{2}=m^{2}-a^{2}$.
The limit $a\to 0$ reduces to the static $q$-metric, while $q\to 0$ reduces to Kerr.
In this section we use the stationary extension only as a phenomenological map from $q$ to low order multipoles; we do not perform a rotating brick-wall/WKB entropy calculation.
The mass and current multipole moments $(M_\ell,J_\ell)$ depart
perturbatively from the Kerr values $M_\ell+iJ_\ell=M(ia)^\ell$.
To lowest orders one finds (e.g., \cite{Toktarbay_2014,Faraji_2022}), namely we have
\begin{align}\label{eq:moments-lowerorder}
    &M_0 =m+q \sigma,\\
    & M_2 = -m^3-3 m^2 q \sigma -m \left(q^2-1\right) \sigma ^2-\frac{1}{3} q \left(q^2-7\right) \sigma ^3,\nonumber \\
    &J_1 = a (m+2 q \sigma ),\nonumber \\
    &J_3=-\frac{a}{3} \left[3 m^3+12 m^2 q \sigma +3 m \left(3 q^2-1\right) \sigma ^2+2 q \left(q^2-4\right) \sigma ^3\right], \nonumber
\end{align}
for the lowest mass and angular momentum multipoles, where $\sigma=\sqrt{m^2-a^2}$ (e.g., \cite{Toktarbay_2014,Faraji_2022}).

We define two dimensionless combinations of these mutipoles that vanish identically for Kerr
\begin{align}\label{eq:Q1massq}
   & Q_1 \equiv \frac{|J_1^2+ M_0 M_2|}{M_0^4} = \frac{2}{3} \left(1-a_*^2\right)^{3/2} |q| + {\cal O}(q^2),\\
    &Q_2 \equiv \frac{|J_1^3+ M_0^2 J_3|}{M^5_0} = \frac{4}{3} |a_*|  \left(1-a_*^2\right)^{3/2} |q| + {\cal O}(q^2),
\end{align}
where $a_* \equiv a/m$ is the dimensionless spin parameter, which ranges between $-1$ and $1$ for Kerr geometry (with $a_* = 0$ for Schwarzschild). We see that violations of the Kerr multipole relations are encoded in $(Q_1,Q_2)$: these Kerr-null combinations vanish for Kerr and acquire a controlled $O(|q|)$ signal in the stationary extension.

\begin{figure}
    \centering    
    \includegraphics[width=\linewidth]{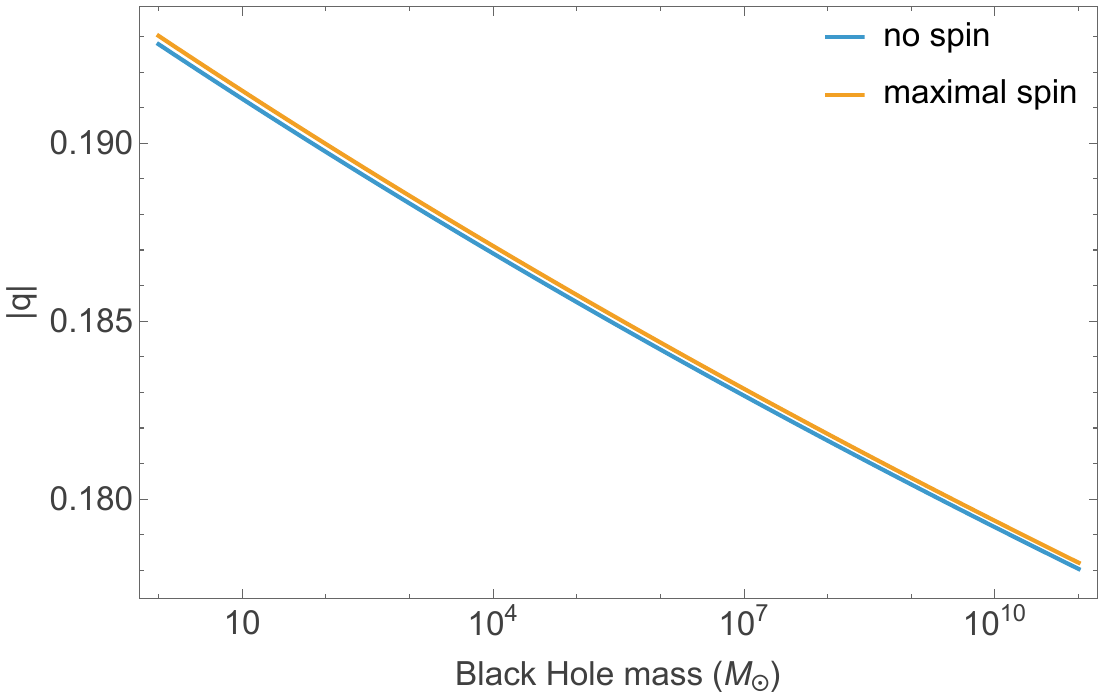}
     \includegraphics[width=\linewidth]{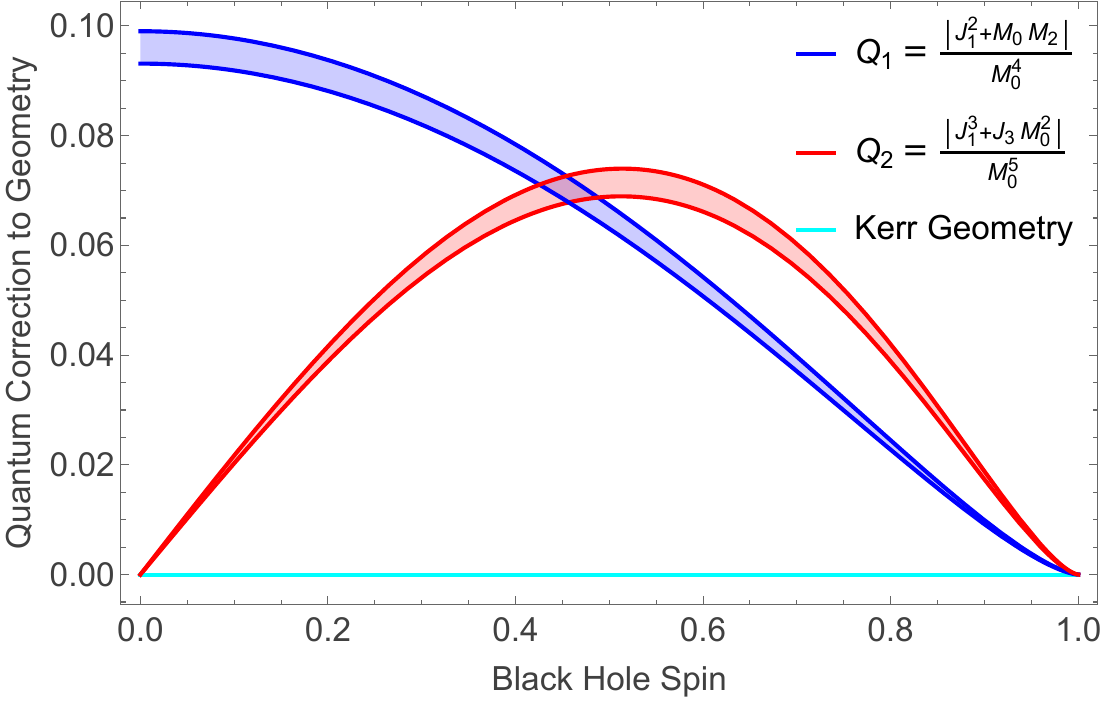}
   \caption{(Top, a) Entropy-motivated estimate of $|q|$ across astrophysical masses obtained from \eqref{eq:accurate-bound}, shown for two macroscopic entropy benchmarks: Schwarzschild ($a_*=0$) and extremal Kerr ($|a_*|=1$), which differ only by an $O(1)$ factor in $S_{\rm BH}$ at fixed mass. (Bottom, b) The corresponding magnitudes of the Kerr-null multipole diagnostics $(Q_1,Q_2)$ in \eqref{eq:Q1massq} as functions of spin, for the same $|q|$ scale.}
    \label{fig:q_mass}
\end{figure}

What value of $q$ should one use at nonzero spin? Our entropy calculation leading to equation \eqref{eq:accurate-bound} is based on the static near-horizon geometry and does not determine a first-principles $q(a_*)$ relation. For the purpose of forecasting the size of Kerr-null multipole diagnostics, we therefore adopt \eqref{eq:accurate-bound} as an entropy-motivated fiducial scale and evaluate it using the macroscopic Bekenstein-Hawking entropy of a Kerr black hole with the same mass and spin, $S_{\rm BH}(m,a_*)$. Since $\ln S_{\rm BH}$ varies only by $O(1)$ (at most $\ln 2$) between Schwarzschild and extremal Kerr at fixed mass, the inferred $|q|$ changes only weakly with spin in this benchmark procedure (Fig.~\ref{fig:q_mass}a).

With this preamble, we use equation \eqref{eq:accurate-bound} only to set an entropy-motivated scale for $|q|$ and translate it into the Kerr-null multipole diagnostics $(Q_1,Q_2)$ through \eqref{eq:Q1massq}. The spin dependence shown in Fig.~\ref{fig:q_mass}a arises solely from the choice of macroscopic entropy benchmark $S_{\rm BH}(m,a_*)$ and should not be interpreted as a derived prediction for $q(a_*)$ in a spinning near-horizon geometry. Fig.~\ref{fig:q_mass}b then shows the corresponding $Q_1$ and $Q_2$ signals. For astrophysical entropies $S_{\rm BH}\sim10^{78}\text{--}10^{97}$ this procedure yields $|q|\sim 0.2$, within the small-$|q|$ regime used throughout and small enough that the exterior remains close to Schwarzschild outside a few gravitational radii; we return to this below.

We note that extending the brick-wall mode count to the stationary metric \eqref{eq:stationary_qmetric}
is nontrivial because, for generic $q\neq 0$, the scalar wave equation need not separate as in Kerr. However, separability is not required to determine the ultraviolet mode density: in the eikonal limit the field modes obey the Hamilton-Jacobi dispersion relation $g^{\mu\nu}p_\mu p_\nu=0$, and in any stationary axisymmetric spacetime the conserved quantities $E=-p_t$ and $L=p_\phi$ allow one to perform WKB counting directly from the invariant phase-space volume with the corotating energy $E-\Omega L$ (where $\Omega$ is the local frame dragging angular velocity). In such a formulation, the regularization mechanism is controlled by the near-horizon scaling of the lapse/redshift and the transverse geometry, while spin enters through $\Omega$ and affects the detailed spectrum and prefactors. We therefore leave the explicit rotating mode count to future work and use the stationary extension here only as a phenomenological map from $q$ to the low order Geroch-Hansen multipoles.

\subsection{Observational signatures}\label{subsec:EHT}

With these predictions in hand, we now outline how two forthcoming experiments could constrain such departures from the Kerr multipole relations.

\subsubsection{Next Generation Event Horizon Telescope (ngEHT)}

At (sub)millimetre wavelengths, the brightness distribution around a supermassive black hole encodes strong-lensing geometry at radii $r\sim \mathcal{O}(m)$. The most robust geometric observables are: (i) the photon ring diameter and its departures from circularity (e.g., axial ratio and quadrupolar distortions) and (ii) the image centroid (ring centre) relative to the dynamical centre. These quantities depend primarily on the spacetime lensing map and are comparatively less sensitive to emissivity once multifrequency coverage and multi-epoch averaging are used.

Quantifying the resulting photon-ring distortions requires ray tracing in the full metric (or its stationary extension); here we restrict to order-of-magnitude sensitivity targets.

The key point is that $q$ perturbs geometry (ring shape/centroid), not just brightness, so it is accessible with geometric/ring-component fits. Under thermal-noise-dominated performance and with calibration systematics under control, ngEHT imaging/visibility fitting at 230/345\,GHz is expected to reach $\mu$as level uncertainties on the ring diameter and a few $\mu$as  on the centroid for bright, stable targets like M87*.  Because these observables are set by null‑geodesic lensing rather than by the details of the emissivity, they are naturally robust to source modeling. In practice, visibility‑domain ring fits and multi‑frequency, multi‑epoch imaging isolate the geometric signal while averaging over variability and scattering, and external constraints on the black hole mass and the angular diameter distance to the source, break the mass distance degeneracy. With the ngEHT’s anticipated microarcsecond level precision, a quadrupolar distortion of the size discussed here would imprint percent‑to‑tens‑of‑percent departures in ring geometry that are well within reach.

Therefore, given that M87* has a light-ring angular scale of order $\sim 40\,\mu$as, ngEHT-class measurements could reach sensitivity to constrain percent-level departures in ring geometry, which would translate into constraints on the Kerr-null deviations discussed here, subject to astrophysical variability and calibration systematics \cite{2023Galax..11...61J,2023Galax..11..107D,2024A&A...681A..79E}.

\subsubsection{Laser Interferometer Space Antenna (LISA)}
\label{subsec:LISA}

Let us now turn our focus to potential observations of extreme-mass-ratio inspirals (EMRIs) by the space-based Laser Interferometer Space Antenna (LISA).
In an EMRI, a compact object of mass $\mu\,\ll\,m$ orbits a massive black hole and slowly inspirals through $\sim 10^4-10^5$ cycles in band e.g., \cite{PhysRevD.95.103012}. The phasing and precession structure of the gravitational waveform encode the multipole structure of the central object. In Kerr geometry, one has $M_{2}=-Ma^{2}$; any deviation of the mass quadrupole moment $M_{2}$ from this relation accumulates a measurable phase shift over the radiation-reaction time. Of course, spin and quadrupole both affect precession, so one needs at least one additional independent multipolar handle to break the quadrupole-spin degeneracy, practically, this comes from combining multiple observables: the two precessions and the frequency‑dependent sideband structure (inclination/polar motion harmonics) and late inspiral behaviour near the last stable orbit. In our language, this is equivalent to measuring one extra Kerr-null combination beyond the quadrupole, so that the fit cannot trade spin against $q$. With LISA-class signal-to-noise, multi-year coherent baselines, and waveform models incorporating precession and higher harmonics, deviations from the Kerr quadrupole can be constrained at the $10^{-4}$ level \cite{PhysRevD.52.5707,PhysRevD.75.042003,PhysRevD.95.103012,2023LRR....26....2A}. Consequently, an $\mathcal{O}(10\%)$ deviation in the relevant multipolar structure would be large compared to the projected statistical precision in idealized EMRI analyses, and could be detectable with LISA provided waveform systematics and parameter degeneracies are under control.

\subsubsection{Ground-Based Gravitational Wave Observatories}

We can also compare our predictions with current bounds from LIGO-Virgo-KAGRA (LVK) ground-based gravitational wave observatories.  
To connect with LVK-style parametrisations, one often introduces a spin-induced quadrupole coefficient, $\kappa$, via $M_2=-\kappa\,\chi^2 M_0^3$, where $\chi:=J_1/M_0^2$. From our definition of $Q_1$ in \eqref{eq:Q1massq}, one has the exact identity
\begin{equation}\label{eq:lizaq}
Q_1=\chi^2\,|\kappa-1|
\qquad\Rightarrow\qquad
|\kappa-1|=\frac{Q_1}{\chi^2},
\end{equation}
Combining \eqref{eq:Q1massq} with $Q_1$ in equation \eqref{eq:lizaq} and using $\chi\simeq |a_*|$ at leading order in small $|q|$ yields the approximate translation
\begin{equation} \label{eq:qkappa}
|q| \simeq \frac{3}{2}\,\frac{a_*^{2}}{(1-a_*^{2})^{3/2}}\,|\kappa-1|\,,
\end{equation}
so a constraint on $|\kappa-1|$ for moderately spinning systems maps directly into a constraint on $|q|$. For example, the bounds on spin and spin-induced quadrupole of GW241011 primary companion are $\chi = 0.78 \pm 0.05$ \cite{LIGOScientific:2025brd}, and $\kappa = 1.1 \pm 0.4$ \cite{Krishnendu:2025rud} (both at 1$\sigma$). Therefore, from \eqref{eq:qkappa}, we find $|q| \lesssim 3$ at 90\% confidence level. Looking ahead, an improvement in SNR by a factor of $\sim 20$ (possible with the next generation of 3G ground-based observatories) can bring our predicted quadrupolar parameter \eqref{eq:qpred} within the detectability range.



\section{Summary and conclusions}
\label{sec:conclusion}

We have shown that a single geometric ingredient (a sufficiently small but nonzero static quadrupolar deformation of the Schwarzschild exterior encoded by the $q$-metric) removes the brick-wall ultraviolet divergence in the canonical entropy of a probe scalar field at any finite asymptotic temperature. In the near-horizon limit the transverse geometry becomes anisotropically scaled, which modifies the effective centrifugal barrier and converts the Schwarzschild logarithmic mode-counting divergence into a convergent power law. Using WKB counting of trapped near-horizon cavity modes (together with the large-$\ell$ angular density approximation), we obtained a closed-form expression for the canonical entropy, which reduces to the standard brick-wall divergence as $q\to 0$.

Matching this canonical entropy $S(q)$ to the macroscopic Bekenstein-Hawking entropy scale $S_{\rm BH}$ (with the thermodynamic inputs specified earlier) yields an entropy-motivated deformation scale. For astrophysical entropies $S_{\rm BH}\sim 10^{78}\text{-}10^{97}$ this gives $|q|\simeq 0.18$-$0.19$, within the small-$|q|$ regime in which the near-horizon expansion is controlled. In other words, the distortion that cures the brick wall is not arbitrarily small: thermodynamic self-consistency fixes its natural scale. In this benchmark, the deformation needed to cure the brick-wall divergence is therefore modest but not parametrically tiny.

Although our entropy calculation is performed in the static geometry, the stationary extension of the $q$-metric provides a convenient phenomenological map from $q$ to measurable multipole deviations: Kerr null combinations such as $Q_1$ and $Q_2$ acquire $O(|q|)$ values. Taking the entropy-motivated scale $|q|\sim 0.2$ as a fiducial target, this corresponds to percent-to-tens-of-percent departures from the Kerr multipole relations for moderate spins, and motivates sensitivity studies with ngEHT strong field lensing observables, EMRI multipole measurements by LISA, and spin-induced quadrupole in 3G ground-based gravitational wave observatories. 

On the theory front, natural next steps include extending the mode count to spinning black holes, treating the angular spectrum beyond the large-$\ell$ approximation, incorporating additional field content, backreaction/renormalization, performing end-to-end ray-tracing and EMRI waveform studies that propagate these deviations into realistic observational inference pipelines.

\acknowledgments
We would like to thank Naritaka Oshita, and Michael Florian Wondrak for useful discussions. This research is supported in part by the University of Waterloo, Natural Sciences and Engineering Research Council of Canada, by the Government of Canada through the Department of Innovation, Science and Economic Development and by the Province of Ontario through the Ministry of Colleges and Universities at Perimeter Institute. 

\bibliographystyle{unsrt}
\bibliography{entropystatic}

\appendix

\section{Klein-Gordon equation components}\label{app:KG}

\paragraph{Time derivative}
\begin{equation}
\frac{1}{\sqrt{-g}}\partial_{t}
 \,\bigl(\sqrt{-g}\,g^{tt}\partial_{t}\phi\bigr)
 =-\frac{1}{f(\rho)}\,\partial_{t}^{2}\phi.
\end{equation}

\paragraph{Radial derivative}
\begin{equation}
\begin{aligned}
\frac{1}{\sqrt{-g}}\,\partial_{\rho}
  \bigl(\sqrt{-g}\,g^{\rho\rho}\partial_{\rho}\phi\bigr)
  &=\frac{1}{4m^{2}\sqrt{f}\sin\theta}\,
    \partial_{\rho}\Bigl(4m^{2}\sqrt{f}\sin\theta\,\partial_{\rho}\phi\Bigr)\\
  &=\partial_{\rho}^{2}\phi
    +\tfrac12\,\frac{f'(\rho)}{f(\rho)}\,\partial_{\rho}\phi.
\end{aligned}
\end{equation}

\paragraph{Angular $\phi$ term}
\begin{equation}
\begin{aligned}
\frac{1}{\sqrt{-g}}\,\partial_{\phi}
 \bigl(\sqrt{-g}\,g^{\phi\phi}\partial_{\phi}\phi\bigr)
&=\frac{1}{4m^{2}\sqrt{f}\,\sin\theta}\,\\
   &\times \partial_{\phi}\,\left(
     4m^{2}\sqrt{f}\sin\theta\,
     \frac{1}{4m^{2}}\left(\frac{\rho}{4m}\right)^{2q}\frac{1}{\sin^{2}\theta}\,
     \partial_{\phi}\phi\right)\\
&=\frac{1}{4m^{2}}\,
   \left(\frac{\rho}{4m}\right)^{2q}\,
   \frac{1}{\sin^{2}\theta}\,\partial_{\phi}^{2}\phi.
\end{aligned}
\end{equation}

\paragraph{Angular $\theta$ term}

\begin{equation}
\begin{aligned}
\frac{1}{\sqrt{-g}}\,\partial_{\theta}
 \bigl(\sqrt{-g}\,g^{\theta\theta}\partial_{\theta}\phi\bigr)
&=\frac{1}{4m^{2}\sqrt{f}\,\sin\theta}\,\\
   &\times \partial_{\theta}\,\left(
     4m^{2}\sqrt{f}\sin\theta\,
     \frac{1}{4m^{2}}\left(\frac{\rho}{4m}\right)^{-2q}
     \partial_{\theta}\phi\right)\\
&=\frac{1}{4m^{2}}\,
   \frac{1}{\sin\theta}\,\partial_{\theta}\,\left(
     \sin\theta\left(\frac{\rho}{4m}\right)^{-2q}\partial_{\theta}\phi\right).
\end{aligned}
\end{equation}
Adding the four pieces gives equation \eqref{eq:box-open} as
\begin{align}
\Box \varphi &\,=
-\frac{1}{f(\rho)}\,\partial_t^{2}\varphi
+\partial_{\rho}^{2}\varphi
+\frac{1}{2}\,\frac{f'(\rho)}{f(\rho)}\,\partial_{\rho}\varphi \\
&+\frac{1}{4m^{2}}\left[
\left(\frac{\rho}{4m}\right)^{-2q}
\frac{1}{\sin\theta}\,\partial_{\theta}\,\left(\sin\theta\,\partial_{\theta}\varphi\right)
+\left(\frac{\rho}{4m}\right)^{2q}
\frac{1}{\sin^{2}\theta}\,\partial_{\phi}^{2}\varphi
\right]\nonumber.
\end{align}


\section{Auxiliary functions for the stationary $q$-metric}
\label{app:stationary_qmetric}

For Equations \eqref{eq:stationary_qmetric_functions} we use \cite{Toktarbay_2014,Faraji_2022}:
\begin{align}
A &= a_+ a_- + b_+ b_-,\qquad
B = a_+^{2}+b_+^{2},\\
C &= (x+1)^{q}\left[x(1-y^{2})(\lambda+\eta)a_+ + y(x^{2}-1)(1-\lambda\eta)b_+\right],\\
a_{\pm} &= (x\pm1)^{q}\left[x(1-\lambda\eta)\pm(1+\lambda\eta)\right],\\
b_{\pm} &= (x\pm1)^{q}\left[y(\lambda+\eta)\mp(\lambda-\eta)\right],\\
\lambda &= \alpha\,(x^{2}-1)^{-q}(x+y)^{2q},\\
\eta &= \alpha\,(x^{2}-1)^{-q}(x-y)^{2q},\\
\alpha &= \frac{\sigma-m}{a}.
\end{align}
The transformation to spheroidal type coordinates $(t,r,\theta,\phi)$ is
$x=(r-m)/\sigma$, $y=\cos\theta$.

\end{document}